\documentclass[twocolumn, nofootinbib, showpacs, amsmath, amssymb]{revtex4}

\usepackage{graphicx}

\begin{document}

\title{ Extra-dimensional generalization of minimum-length deformed QM/QFT\\ and some of its phenomenological consequences}

\author{Michael~Maziashvili }
\email{maziashvili@iliauni.edu.ge} \affiliation{School of Natural Sciences and Engineering, Ilia State University,\\ 3/5 Cholokashvili Ave., Tbilisi 0162, Georgia}

\begin{abstract}

In contrast to the 3D case, different approaches for deriving the gravitational corrections to the Heisenberg uncertainty relation do not lead to the unique result whereas additional spatial dimensions are present in the theory. We suggest to take logarithmic corrections to the black hole entropy, which has recently been proved both in string theory and loop quantum gravity to persist in presence of additional spatial dimensions, as a point of entry for identifying the modified Heisenberg-Weyl algebra. We then use a particular Hilbert space representation for such a quantum mechanics to construct the correspondingly modified field theory and address some phenomenological issues following from it. Some subtleties arising at the second quantization level are clearly pointed out. Solving the field operator to the first order in deformation parameter and defining the modified wave function for a free particle, we discuss the possible phenomenological implications for the black hole evaporation. Putting aside modifications arising at the second quantization level, we address the corrections to the gravitational potential due to modified propagator (back reaction on gravity) and see that correspondingly modified Schwarzschild-Tangherlini space-time shows up the disappearance of the horizon and vanishing of surface gravity when black hole mass approaches the quantum gravity scale. This result points out to the existence of zero-temperature black hole remnants.

\end{abstract}




\maketitle

\subsection{Introduction}

Quantum theory based on a so called minimum-length deformed uncertainty relation is endowed basically with two new features: 1) the modified dispersion relation and 2) the cutoff on the "standard" momentum \cite{Maziashvili:2012zr}. Remarkably enough, this sort of uncertainty relation in 3D can uniquely be reached from various {\tt Gedankenexperimente}, which in higher dimensions lead to the ambiguous result. As a guiding principle for identifying the minimum-length deformed quantum theory in higher dimensions, we suggest to use logarithmic corrections to the black hole (BH) entropy. The computations made in recent years in the framework of string theory \cite{Sen:2012dw} and loop quantum gravity \cite{Bodendorfer:2013sja} demonstrate that the logarithmic corrections to the BH entropy are universal in arbitrary space-time dimensions $\geq 4$. Taking this fact into account, first we consider a few examples of deriving logarithmic corrections to the BH entropy in 3D case by using modified uncertainty relation (MUR). We shall along the way comment on the misleading issues concerning the immediate (heuristic) application of the MUR to BH radiation. Simple physical picture behind this consideration allows one to guess higher-dimensional generalization of minimum-length deformed quantum mechanics (QM). The deformed QM derived this way disagrees with the result that follows from the well known arguments \cite{Maggiore:1993rv, Scardigli:1999jh, Adler:1999bu} (and some other closely related arguments \cite{Maziashvili:2011dx}) for estimating the gravitational corrections to the uncertainty relation. The rest of the paper is devoted to the discussion of quantum field theory (QFT) in view of the deformed quantization both at first and second quantization levels. Phenomenological implications of this study for the black hole physics is explored.

\subsection{Identifying the Planck-length deformed QM with the use of BH entropy corrections}\label{Abschnitt1}

\subsubsection{From MUR to BH entropy corrections: 3D case}\label{Unterabschnitt1}

We start by pointing out that in 3D the MUR being applied to the BH radiation either in an immediate heuristic way or first finding the corresponding Hilbert space representation and then using it for field theory both at first and second quantization levels, uniquely leads to the logarithmic corrections to the BH entropy. The system of units used throughout is: $\hbar=c=1$. The corrections to the BH radiation obtained in a heuristic manner in \cite{Adler:2001vs} can be viewed as a result of modified dispersion relation. Namely, in applying the MUR    
 
\begin{eqnarray}\label{genunrel}
\delta X\delta P \simeq 1 \,+\, \beta l_P^2\delta P^2~, \end{eqnarray} to the BH emission, one assumes that $\delta x$ is set by the horizon radius and consequently the characteristic momentum for the emitted particle is estimated as \begin{eqnarray}\label{momentumofescaping} \delta P \simeq \frac{\delta X \,-\, \sqrt{\delta X^2 \,-\, 4\beta l_P^2}}{2\beta l_P^2} ~.
\end{eqnarray} So, in Eq.\eqref{momentumofescaping} $\delta P$ is understood as a momentum of particle escaping from the BH in the case when correction term for the uncertainty relation is assumed \eqref{genunrel} while $\delta X^{-1}$ defines the momentum in the standard case. Adopting the notations: $P \equiv \delta P$, and $p\equiv \delta X^{-1}$, the Eq.\eqref{genunrel} can be read off as a modified dispersion relation

\begin{eqnarray}\label{moduncererti}
P \,=\, \frac{p^{-1} \,-\,\sqrt{p^{-2} \,-\, 4\beta l_P^2}}{2\beta l_P^2} \,=\, \frac{1 \,-\, \sqrt{1 \,-\, 4\beta l_P^2 \,p^2}}{2\beta l_P^2 \, p}~.
\end{eqnarray} This relation is qualitatively different from that one that follows from the Hilbert space representation of the uncertainty relation \eqref{genunrel} \cite{Kempf:1996nk}  

\begin{eqnarray}\label{moduncori}
P \,=\, \frac{p}{1 \,-\, \beta l_P^2 \,p^2} ~, 
\end{eqnarray} but nevertheless in the low energy regime $p \ll E_P$ both of the Eqs.(\ref{moduncererti}, \ref{moduncori}) look the same

\begin{eqnarray}\label{disptanafardobaerti}
P \,=\, p \,+\, \beta l_P^2 p^3 \,+\, O\left(l_P^4p^5\right) ~.
\end{eqnarray} Let us notice that in view of Eq.\eqref{moduncererti} one has to admit that the momentum $P$ is bounded from above by the Planck energy because when $p > E_P/2\sqrt{\beta}$ it becomes merely a complex quantity. In light of Eq.\eqref{moduncori} the momentum $P$ is not UV bounded, but again $p$ should be restricted to the interval $0 \leq  p \leq E_P/\sqrt{\beta}$ as it is enough to cover the whole momentum space: $0 \leq P < \infty$. In both cases $P$ is understood as a physical momentum that might be used for estimating energy

\[ \varepsilon \,= \sqrt{\, P^2 \,+\, m^2}~, \] while $\mathbf{p}$ is merely a new coordinate in momentum space, which (quantum mechanically) is related to the translations: $\widehat{\mathbf{p}} = -i\nabla$.

Yet another dispersion relation that comes by applying the deformed quantization with respect to Eq.\eqref{genunrel} to the field theory looks like \cite{Berger:2010pj}  

\begin{eqnarray}\label{moduncsami}
\varepsilon \,=\, \sqrt{p^2 \,+\, m^2} \,+\, \beta l_P^2 \, \frac{p^2 \,+\, m^2}{l_\star}~,
\end{eqnarray} where $l_* = E^{-1}$ is set by the characteristic energy scale of the problem under consideration. In the context of BH emission it is just the temperature of the emission. So that the Eq.\eqref{moduncsami} can be put in the form 

  \begin{eqnarray}\label{disptanafardobaori}
\varepsilon \,=\, \sqrt{p^2 \,+\, m^2} \,+\, \beta l_P^2 \, \left(p^2 \,+\, m^2\right)^{3/2}~,
\end{eqnarray}

The effect of Eqs.(\ref{disptanafardobaerti}, \ref{disptanafardobaori}) on the BH emission temperature is that it gets increased as 

\[ T \,\rightarrow \, T \,+\, \beta l_P^2T^3 ~,  \] and correspondingly \[ dS \,=\, \frac{dM}{T} \, \rightarrow \, \frac{dM}{T} \,-\, \beta TdM ~. \] By taking into account that $T \propto r_g^{-1} \simeq 1/l_P^2 M$, to the first order in $\beta$ one finds a logarithmic correction to the entropy

\[ S \,=\, \pi \left(\frac{r_g}{l_P} \right)^2 \,-\, \gamma \ln \left(\frac{r_g}{l_P}\right)~. \] It is worth noticing that the $l_P^4p^5$ term in Eq.\eqref{disptanafardobaerti} results in the inverse area corrections to the entropy.

\subsubsection{From BH entropy corrections to MUR: D $>$ 3 }

The fact that in higher dimensions one also expects the logarithmic corrections to the BH entropy \cite{Sen:2012dw, Bodendorfer:2013sja} can be used to guess the corresponding higher-dimensional generalization of the minimum-length deformed QM. In higher-dimensional case the gravitational radius (which determines the Hawking temperature: $T\propto r_g^{-1}$) looks like: $r_g \simeq \left(l_F^{2+n}M\right)^{1/(1+n)}$, where $n$ denotes the number of extra dimensions and $l_F$ stands for higher-dimensional scale of gravity: $l_F^{2+n} \equiv G_N$. Previous discussion makes it clear that the modified dispersion relation 

\begin{eqnarray}
P \,=\, p \,+\, \beta l_F^{\alpha}  p^{\alpha +1} \,+\, \ldots ~,
\end{eqnarray} will reproduce logarithmic correction to the BH entropy if $\alpha = 2+n$. It suggests the minimum-length deformed QM of the form \begin{equation}\label{hdmcomr} \left[ \widehat{X},\, \widehat{P}  \right]  \,=\, i \left(1 \, +\, \beta \, l_F^{2+n} \widehat{P}\,^{2+n} \right) ~,  \end{equation} which indeed implies the existence of the minimum position uncertainty of the order of \cite{Maslowski:2012aj} 

\[ \delta X \, \simeq \,  \left[\int\limits_0^{\infty} \frac{dP}{1 \, +\, \beta \, l_F^{2+n} P\,^{2+n}} \right]^{-1} \,=\, \frac{\beta^\frac{1}{2+n}l_F}{\int\limits_0^{\infty} \frac{dq}{1 \, +\, q\,^{2+n}} }~.  \] As in the 3D case \cite{Kempf:1996nk}, the algebra \eqref{hdmcomr} maybe written in a somewhat generic form \cite{Maziashvili:2012dd}

\begin{eqnarray}\label{multdimcr}&& \left[\widehat{X}_i,\, \widehat{X}_j \right] \,=\, 0~,~ \left[\widehat{P}_i,\, \widehat{P}_j \right] \,=\, 0 ~, \nonumber \\&& 
 \left[\widehat{X}_i,\, \widehat{P}_j \right] \,=\, i\left\{  \Xi\left(\widehat{P}^2\right) \delta_{ij} \,+\, \Theta \left(\widehat{P}^2\right)\widehat{P}_i\widehat{P}_j \right\}~,
\end{eqnarray} where the simplest {\tt ansatz} for $\Theta$ is understood

\begin{equation} \Theta \left(\widehat{P}^2\right) \,=\, 2\beta \,l_F^{2+n} \widehat{P}^{n}  ~.\nonumber \end{equation} The Hilbert space representation of Eq.\eqref{multdimcr} can be constructed in terms of the standard $\widehat{\mathbf{x}},\, \widehat{\mathbf{p}}$ operators as \cite{Maziashvili:2012dd}

\begin{eqnarray}
\widehat{X}_j \,=\, \widehat{x}_j~,~~ \widehat{P}_j \,=\, \frac{\widehat{p}_j}{\left( 1\,-\, \frac{2\beta (1+n)}{2+n} \, l_F^{2+n}\widehat{p}\,^{2+n} \right)^\frac{1}{1+n}}~,
\end{eqnarray} or in the eigen-representation of the $\widehat{\mathbf{p}}$ operator

\begin{eqnarray}
\widehat{X}_j \,=\, i \, \frac{\partial}{\partial p_j}~,~~ \widehat{P}_j \,=\, \frac{p_j}{\left( 1\,-\, \frac{2\beta (1+n)}{2+n} \, l_F^{2+n}p\,^{2+n} \right)^\frac{1}{1+n}}~,~ \end{eqnarray} with scalar product containing a cut-off on $p$

\begin{eqnarray} \langle \psi_1 |\psi_2 \rangle \,=\, \int\limits_{p^{2+n}<(2+n)/2\beta(1+n)l_F^{2+n}} d^{3+n}p \, \psi^*_1(\mathbf{p})\psi_2(\mathbf{p})~. ~~
\end{eqnarray} Let us notice that this construction is a straightforward generalization of the 3D picture described in \cite{Kempf:1996nk}. Here the cutoff $p^{2+n}<(2+n)/2\beta(1+n)l_F^{2+n}$ has the same meaning as in Eq.\eqref{moduncori}. Now the analog of Eq.\eqref{moduncori} takes the form 

\begin{eqnarray}
P \,=\, \frac{p}{\left( 1\,-\, \frac{2\beta (1+n)}{2+n} \, l_F^{2+n}p\,^{2+n} \right)^\frac{1}{1+n}} \,=\, \nonumber \\ p \,+\, \frac{2\beta l_F^{2+n}p^{3+n}}{2+n}  \,+\, \frac{2\beta^2 l_F^{4+2n}p^{5+2n}}{2+n} \,+\, \ldots  ~,
\label{highdimmomentumrel} \end{eqnarray} where the $ l_F^{2+n}$ term reproduces the logarithmic correction to the BH entropy while the $ l_F^{4+2n}$ term is responsible for the inverse area correction.

\subsubsection{Comparing with the result following from {\tt Gedankenexperimente} usually used in 3D }

For the sake of comparison, here we briefly discuss the MUR that follows from the {\tt Gedankenexperimente} taking into account the gravitational effect in particle's position measurement or some simple dimensional arguments. It is worth noticing that in 3D one has an unique picture for various approaches. Let us first look at the dimensional arguments for the gravitational corrections to the Heisenberg uncertainty relation that results in a lower bound on position uncertainty. For our purposes it will be convenient to choose the system of units: $c=1$; that is, $[\hbar]=$g$\cdot$cm,\,$[\mathbb{G}_N]=$cm$^{n+1}/$g. Just on the dimensional grounds, one can write somewhat generic expression for MUR

\begin{eqnarray}\label{gur1}\delta X \delta P \, &\geq & \, \frac{\hbar}{2} \,+\, \beta \, \hbar^{(\alpha -1)/\alpha} \mathbb{G}_N^{1/\alpha(n+1)}\delta P^{(n+2)/\alpha(n+1)} ~,~~~~~  \end{eqnarray} where $\beta$ is a numerical factor of order unity. In order to have a lower bound on position uncertainty, one should require

\begin{eqnarray}  \alpha \, \leq \, \frac{n+2}{n+1}~. \nonumber \end{eqnarray} On the other hand, to allow the limit $\hbar \rightarrow 0$, one has to require $\alpha \geq 1$. It is immediate to see that if one picks out the value: $\alpha = 1$ then the correction term in Eq.\eqref{gur1} does not depend on $\hbar$ and therefore survives even in the limit $\hbar \rightarrow 0$. By taking this specific choice one arrives at the equation

\begin{equation}\label{gurhigher} \delta X \delta P  \,\geq \,
  \frac{\hbar}{2} \,+\, \beta \, \mathbb{G}_N^\frac{1}{n+1}\delta P^\frac{n+2}{n+1}~.  \end{equation} From now on we will again adopt the system of units $\hbar = c =1$ and discuss the correction term in Eq.\eqref{gurhigher} as a result of certain gravitational effects.

In the case $\delta P \ll \mathbb{G}_N^{-1/(2+n)}$ the correction term in Eq.\eqref{gurhigher} can be considered as a result of the gravitational extension of the wave-packet localization width as compared to the Minkowskian background \cite{Maziashvili:2011dx}. Yet, the correction term in Eq.\eqref{gurhigher} makes sense even for $\delta P \gtrsim \mathbb{G}_N^{-1/(2+n)}$. In this case it is motivated by the fact that in a high center of mass energy scattering, $\sqrt{s} \gtrsim \mathbb{G}_N^{-1/(2+n)}$, the production of BH dominates all perturbative processes \cite{'tHooft:1987rb, Antoniadis:1998ig, Giddings:2001bu, Kaloper:2007pb}, thus limiting the ability to probe short distances. (It is important to notice that at high energies, $\sqrt{s} \gg \mathbb{G}_N^{-1/(2+n)}$, the BH production is increasingly a long-distance, semi-classical process). To make the point clearer, the refined measurement of particle's position requires large energy transfer during a scattering process used for the measurement. But when the gravitational radius associated with this energy transfer $\sim \left(\mathbb{G}_N\sqrt{s}\,\right)^{1/(1+n)}$ becomes grater than the impact parameter, the BH will form and what one can say about the particle's position is that it was somewhere within the region $\sim \left(\mathbb{G}_N\sqrt{s}\,\right)^{1/(1+n)}$. The gravitational radius of the BH formed in the scattering process grows with energy as $r_g \simeq \left(\mathbb{G}_N\sqrt{s}\,\right)^{1/(1+n)}$ determining therefore high energy behavior of the position uncertainty.

To summarize, in $D > 3$ the deformed QM given by Eq.\eqref{hdmcomr} might be favoured over the suggestion made in \cite{Scardigli:2003kr} as it allows to reproduce the logarithmic and inverse area corrections to the BH entropy, which, in its turn, seems to be a sound result irrespective to the number of dimensions \cite{Sen:2012dw, Bodendorfer:2013sja}. Let us notice that the MUR closely related to the deformed QM \eqref{hdmcomr} was suggested in a somewhat different context in \cite{Aurilia:2002aw}.

\subsection{Free field in $3+n$ dimensions }

In this section we just recapitulate some textbook material \cite{LL} to prepare a background for discussing the minimum-length deformed QFT. Let us consider a neutral scalar field  $\varPhi$ in a finite volume $l^{3+n}$

\[ H = \int\limits_{l^{3+n}} d^{3+n}x \,\, \frac{1}{2} \left[ \varPi^2 + \partial_{\mathbf{x}} \varPhi\partial_{\mathbf{x}} \varPhi + m^2  \varPhi^2 \right]~, \] where $\varPi = \dot{\varPhi}$. After using the Fourier expansion for $\varPi$ and $\varPhi$ \begin{eqnarray} \varPhi(\mathbf{x}) = \frac{1}{l^{3+n}} \sum\limits_{\mathbf{p}_n} \varphi (\mathbf{p}_n)\, e^{i\mathbf{p}_n\mathbf{x}}~, \nonumber \\  \varPi(\mathbf{x}) = \frac{1}{l^{3+n}} \sum\limits_{\mathbf{p}_n} \pi (\mathbf{p}_n)\, e^{i\mathbf{p}_n\mathbf{x}}~, \nonumber \end{eqnarray} the Hamiltonian takes the form

\[ H = \frac{1}{2 l^{3+n}} \sum\limits_{\mathbf{p}_n} \left[ \pi(\mathbf{p}_n)\pi^+(\mathbf{p}_n) + (\mathbf{p}_n^2 + m^2)\varphi(\mathbf{p}_n)\varphi^+(\mathbf{p}_n) \right] ~. \] The quantization conditions 
\begin{eqnarray} && \left[\varPhi(\mathbf{x}),\, \varPi(\mathbf{y}) \right] = i\delta (\mathbf{x} - \mathbf{y})~,~~\left[\varPhi(\mathbf{x}),\, \varPhi(\mathbf{y}) \right] =0 ~,  \nonumber \\ && \left[\varPi(\mathbf{x}),\, \varPi(\mathbf{y}) \right] = 0~,  \nonumber\end{eqnarray} for the Fourier modes imply

\begin{eqnarray} && \left[\varphi(\mathbf{p}_n),\, \pi(\mathbf{p}_m)\right] = i\,l^{3+n}\,\delta_{-\mathbf{p}_n\mathbf{p}_m}~,~~\left[\varphi(\mathbf{p}_n),\, \varphi(\mathbf{p}_m)\right] = 0 ~,\nonumber \\ && \left[\pi(\mathbf{p}_n),\, \pi(\mathbf{p}_m)\right] = 0~. \nonumber \end{eqnarray} Defining 

\begin{eqnarray}&& a(\mathbf{p}_n) = \frac{1}{\sqrt{2\varepsilon_{\mathbf{p}_n}}} \left[\varepsilon_{\mathbf{p}_n}\varphi(\mathbf{p}_n) + i  \pi(\mathbf{p}_n) \right]~, \nonumber \\ && a^+(\mathbf{p}_n) = \frac{1}{\sqrt{2\varepsilon_{\mathbf{p}_n}}} \left[\varepsilon_{\mathbf{p}_n}\varphi(-\mathbf{p}_n) - i  \pi(-\mathbf{p}_n) \right]~,\nonumber \end{eqnarray} where $\varepsilon_{\mathbf{p}_n} = \sqrt{\mathbf{p}_n^2 + m^2}$, one finds 

\begin{eqnarray} && \left[a(\mathbf{p}_n),\,a^+(\mathbf{p}_m) \right] = l^{3+n}\delta_{\mathbf{p}_n\mathbf{p}_m}~,~~ \left[a(\mathbf{p}_n),\,a(\mathbf{p}_m) \right] =0~, \nonumber \\ && \left[a^+(\mathbf{p}_n),\,a^+(\mathbf{p}_m) \right] =0~. \nonumber \label{standardcreanni}\end{eqnarray} So, the field and momentum operators take the form 

\begin{eqnarray} \varPhi(\mathbf{x}) &=& \frac{1}{l^{3+n}} \sum\limits_{\mathbf{p}_n} \frac{1}{\sqrt{2\varepsilon_{\mathbf{p}_n}}} \left[ a(\mathbf{p}_n)e^{i\mathbf{p}_n\mathbf{x}} + a^+(\mathbf{p}_n) e^{-i\mathbf{p}_n\mathbf{x}} \right]  \nonumber ~, \nonumber \\ 
\varPi(\mathbf{x}) &=&  \frac{i}{l^{3+n}} \sum\limits_{\mathbf{p}_n}   \sqrt{\frac{\varepsilon_{\mathbf{p}_n}}{2}}  \left[   a^+(\mathbf{p}_n)e^{-i\mathbf{p}_n\mathbf{x}} -  a(\mathbf{p}_n)e^{i\mathbf{p}_n\mathbf{x}} \right]  \nonumber ~, \end{eqnarray} and the Hamiltonian reduces to  

\begin{equation} H = \frac{1}{2l^{3+n}}   \sum\limits_{\mathbf{p}_n}  \varepsilon_{\mathbf{p}_n} \left[ a^+(\mathbf{p}_n)a(\mathbf{p}_n) + a(\mathbf{p}_n)a^+(\mathbf{p}_n) \right] ~. \label{hamintermsaadagger} \nonumber\end{equation} Introducing real variables 

\begin{eqnarray} && Q_{\mathbf{p}_n} =\, \frac{1}{\sqrt{2\mu l^{3+n}\varepsilon_{\mathbf{p}_n}}} \left[ a(\mathbf{p}_n) + a^+(\mathbf{p}_n)\right]~,\nonumber \\ && P_{\mathbf{p}_n} = \, i\sqrt{\frac{\mu\varepsilon_{\mathbf{p}_n}}{2l^{3+n} }} \left[a^+(\mathbf{p}_n) - a(\mathbf{p}_n)\right] ~,\nonumber \end{eqnarray} the Hamiltonian splits into a sum of independent one-dimensional oscillators

\begin{equation} H =   \sum\limits_{\mathbf{p}_n} \left(\frac{ P_{\mathbf{p}_n}^2}{2\mu} + \frac{\mu\varepsilon_{\mathbf{p}_n}^2 Q_{\mathbf{p}_n}^2}{2} \right)~.\label{oscillsum}\end{equation}

We explicitly introduced an energy scale $\mu$ in order the variables $Q_{\mathbf{p}_n}, P_{\mathbf{p}_n}$ to have the natural dimensions: $[Q_{\mathbf{p}_n}]=\,$cm and $[P_{\mathbf{p}_n}]=\,$cm$^{-1}$. So far the parameter $\mu$ is entirely arbitrary. The basic idea behind explicitly introducing this parameter is a characteristic feature of the minimum-length deformed quantization that it engenders a mass dependence of the oscillator energy spectrum \cite{Kempf:1994su, Kempf:1996fz} while the standard quantization scheme is free of this feature.  So that the quantization of the field, suitably altered to respect the effects of a minimal length, necessarily involves some characteristic energy scale $\mu$ in the vein of an effective QFT. For identifying the energy scale $\mu$, one may keep in mind that in view of Eq.\eqref{hdmcomr} the deviation from the standard quantization becomes appreciable at high energies. Therefore it naturally suggests the identification of $\mu$ with the characteristic energy scale of the problem under consideration. This sort of reasoning is completely in the spirit of an effective QFT \cite{Berger:2010pj}. 

The Heisenberg equation of motion reads 

\[ \dot{a}(\mathbf{p}_n) = i \left[H, a(\mathbf{p}_n)\right] = - i \varepsilon_{\mathbf{p}_n}a(\mathbf{p}_n) ~, \] which can be solved as 

\[a(t,\,\mathbf{p}_n) = a(t = 0,\,\mathbf{p}_n) e^{-i\varepsilon_{\mathbf{p}_n} t} ~.\] The field and momentum operators take the form

\begin{widetext}
\begin{eqnarray} \varPhi(t,\,\mathbf{x})  & = & \frac{1}{l^{3+n}} \sum\limits_{\mathbf{p}_n} \frac{1}{\sqrt{2\varepsilon_{\mathbf{p}_n}}} \left[ a(0,\,\mathbf{p}_n) e^{i(\mathbf{p}_n\mathbf{x} - \varepsilon_{\mathbf{p}_n} t)} + a^+(0,\,\mathbf{p}_n) e^{-i(\mathbf{p}_n\mathbf{x} - \varepsilon_{\mathbf{p}_n} t)} \right]  ~, \nonumber\\
\varPi(t,\,\mathbf{x}) &=&  \frac{i}{l^{3+n}} \sum\limits_{\mathbf{p}_n}   \sqrt{\frac{\varepsilon_{\mathbf{p}_n}}{2}}  \left[   a^+(0,\,\mathbf{p}_n) e^{-i(\mathbf{p}_n\mathbf{x} - \varepsilon_{\mathbf{p}_n} t)} -  a(0,\,\mathbf{p}_n) e^{i(\mathbf{p}_n\mathbf{x} - \varepsilon_{\mathbf{p}_n} t)} \right]  \nonumber 
 ~. \end{eqnarray} 

\end{widetext} Then, we write $a(\mathbf{p}_n)$ for $a(0,\,\mathbf{p}_n)$, and similarly $a^+(\mathbf{p}_n)$ for $a^+(0,\,\mathbf{p}_n)$ in field theory and call these quantities the annihilation and creation operators, respectively.

\subsection{ Minimum-length deformed QFT}

As long as we are restricting ourselves to the leading order corrections due to minimum-length deformed quantum theory, the corrections arising at the first and second quantization levels do not interfere and can be considered separately.

\subsubsection{Corrections arising at the first quantization level}

The modified field theory 

\begin{eqnarray}\label{wirkung}
\mathcal{W}[\varPhi] = -\int d^{4+n}x \left(\varPhi\partial_t^2\varPhi + \varPhi \widehat{\mathbf{P}}^2\varPhi  +m^2\varPhi^2 \right) ~,
\end{eqnarray} leads to the equation of motion 

\begin{equation}
\partial_t^2\varPhi + \widehat{\mathbf{P}}^2\varPhi  +m^2\varPhi = 0 ~,
\end{equation} which in its turn admits the plane wave solution $\sim \exp(i\mathbf{p}\mathbf{x})$ with a modified dispersion relation 

\begin{equation}
\varepsilon^2 \,=\, \mathbf{P}^2 \,+\, m^2 \,=\,\frac{p^2}{\left( 1\,-\, \frac{2\beta (1+n)l_F^{2+n}p^{2+n}}{2+n}  \right)^\frac{2}{1+n}} \,+\, m^2~.
\end{equation} This dispersion relation implies the superluminal motion; namely, taking $m=0$ one finds 

\begin{equation}
\frac{d\varepsilon}{dp} \,=\, \frac{2\,+\, n \,+\, 2\beta l_F^{2+n}p^{2+n}}{2\,+\, n \,-\, 2\beta (1+n) l_F^{2+n}p^{2+n}} \, > \, 1~.
\end{equation}



\subsubsection{Corrections arising at the second quantization level}

The corrections at the second quantization level are obtained by quantizing the field Hamiltonian with respect to the Eq.\eqref{hdmcomr}. In effect the appearance of energy scale $\mu$ besides $l_F^{-1}$ lends the possibility for introducing a dimensionless parameter $\left(\mu l_F \right)^{2+n}$ that measures the deviation from the standard picture in accordance with the Eq.\eqref{hdmcomr}.     
For each oscillator entering the Eq.\eqref{oscillsum} now we have \eqref{hdmcomr}  
\begin{equation} \left[Q_{\mathbf{p}_n},\,P_{\mathbf{p}_m}\right] = i\delta_{\mathbf{p}_n\mathbf{p}_m} \left(1 + \ss P^{2+n}\right)~,\label{mlqm}\end{equation} where we have used the notation 

\begin{equation}\label{ersteBezeichnung}
\beta l_F^{2+n} \, \equiv  \, \ss ~.
\end{equation} To the first order in $\beta$, from Eq.\eqref{hdmcomr} one finds 

\begin{equation}
\widehat{X} \,=\, \widehat{x}~, ~~\widehat{P} \,=\, \widehat{p} \,+\, \frac{\ss \, \widehat{p\,}^{3+n}}{3+n} \,+\, O\left(\ss^2\right)~.
\end{equation} Therefore, the Hamiltonian 

\[ H \,=\, \frac{P^2}{2\mu} \,+\, \frac{\mu\varepsilon^2 Q^2}{2} ~,\] to the first order in $\beta$ takes the form 

\begin{eqnarray}  H  &=& \frac{p^2}{2\mu} \,+\, \frac{\mu\varepsilon^2 q^2}{2} \,+\, \frac{\ss\, p^{4+n}}{\mu (3+n)}  \nonumber\\ &=& \varepsilon \left( b^+b + \frac{1}{2} \right) \,+\, \frac{\ss i^{4+n}}{\mu (3+n)}\left(\frac{\mu\varepsilon}{2}\right)^\frac{4+n}{2}(b^+ - b)^{4+n} ~, \nonumber \end{eqnarray}
where \[ b = \frac{1}{\sqrt{2\mu\varepsilon}} \left( \mu\varepsilon q + i p \right) ~,~~~~ b^+ = \frac{1}{\sqrt{2\mu\varepsilon}} \left( \mu\varepsilon  q - i p \right) ~.\] Using this Hamiltonian, from the Heisenberg equation $\dot{b} = i \left[H,\,b\right]$ one finds 

\begin{equation} \dot{b} \,=\, -i\varepsilon b \,-\, \frac{(4+n)i^{5+n}\ss}{\mu(3+n)}\left(\frac{\mu\varepsilon}{2}\right)^\frac{4+n}{2}(b^+ - b)^{3+n} ~.\label{heqlin}\end{equation} Writing the operator $b$ to the first order in $\ss$ in the form 

\[ b = f + \ss g~,\] then Eq.\eqref{heqlin} takes the form 

\begin{equation} \dot{f} + \ss \dot{g} = -i\varepsilon (f + \ss g) \,-\, \ss \aleph \, (f^+ - f)^{3+n}~,\label{heqfirst}\end{equation} where we have used the notation 

\begin{equation}
\aleph \,=\, \frac{(4+n)i^{1+n}}{\mu(3+n)}\left(\frac{\mu\varepsilon}{2}\right)^\frac{4+n}{2} ~.
\end{equation} Equating the coefficients of like powers of $\ss$ from Eq.\eqref{heqfirst} one finds

\begin{equation} \dot{f} = - i\varepsilon f~,~~ \dot{g} = -i\varepsilon g \,-\, \aleph \, (f^+ - f)^{3+n} ~,\nonumber\end{equation} which admits the following analytic solution 

\begin{eqnarray} && f(t) = f(0) e^{-i\varepsilon t}~, \nonumber \\ && \dot{g}  = -i\varepsilon g \,-\, \aleph \, \left[ f^+(0) e^{i\varepsilon t} - f(0) e^{-i\varepsilon t}\right]^{3+n} ~, \nonumber \\ \label{heisenbergssol} && g(t) =  \\ && e^{-i\varepsilon t } \left[ g(0) - \aleph \int\limits_0^t d\tau\, e^{i\varepsilon \tau} \left\{ f^+(0) e^{i\varepsilon \tau} - f(0) e^{-i\varepsilon \tau}\right\}^{3+n} \right]~. \nonumber \end{eqnarray} Using Eq.\eqref{heisenbergssol} to the first order in $\ss$ one can write 

\begin{eqnarray} b(t) \,=\, b(0)e^{-i\varepsilon t} \,\,- ~~~~~~~~~~~~~~~~~~~~~~~~~~~~~~~~~~~~\nonumber \\ \ss \aleph \,e^{-i\varepsilon t} \int\limits_0^t d\tau\, e^{i\varepsilon \tau} \left\{ b^+(0) e^{i\varepsilon \tau} - b(0) e^{-i\varepsilon \tau}\right\}^{3+n} ~.\nonumber\end{eqnarray} Thus, the corrected field operator takes the form 

\begin{widetext}

\begin{eqnarray}\label{Feldquantisierung} \varPhi(t,\,\mathbf{x}) \, =\, \frac{1}{l^{3+n}} \sum\limits_{\mathbf{p}_n} \frac{1}{\sqrt{2\varepsilon_{\mathbf{p}_n}}} \left[ \left(b(\mathbf{p}_n) \,-\, \ss \aleph \, \int\limits_0^t d\tau\, e^{i\varepsilon_{\mathbf{p}_n} \tau} \left[ b^+(\mathbf{p}_n) e^{i\varepsilon_{\mathbf{p}_n} \tau} - b(\mathbf{p}_n) e^{-i\varepsilon_{\mathbf{p}_n} \tau}\right]^{3+n} \right)e^{i(\mathbf{p}_n\mathbf{x} - \varepsilon_{\mathbf{p}_n} t)} \right.\nonumber\\ \left. + \left(b^+(\mathbf{p}_n) \,-\, \ss \aleph^* \, \int\limits_0^t d\tau\, e^{-i\varepsilon_{\mathbf{p}_n} \tau} \left[ b(\mathbf{p}_n) e^{-i\varepsilon_{\mathbf{p}_n} \tau} - b^+(\mathbf{p}_n) e^{i\varepsilon_{\mathbf{p}_n} \tau}\right]^{3+n} \right)e^{-i(\mathbf{p}_n\mathbf{x} - \varepsilon_{\mathbf{p}_n} t)} \right]   ~. \end{eqnarray}

\end{widetext} Keeping in mind that at a fundamental level the notion of particle (quantum) comes from the quantized field, we define free particle wave function by means of the matrix element $\langle 0| \varPhi(t,\,\mathbf{x}) | \mathbf{p}_i \rangle$, which in the standard case gives just the de Broglie wave. Following this definition and using Eq.\eqref{Feldquantisierung} we estimate corrections to the free particle wave function due to minimum-length deformed QM to the first order in the deformation parameter $\ss$. One immediately sees that if $n$ is odd, then the matrix element $\langle 0| \varPhi(t,\,\mathbf{x}) | \mathbf{p}_i \rangle \propto e^{i(\mathbf{p}_i\mathbf{x} - \varepsilon_{\mathbf{p}_i} t)}$. Let us assume $n$ is an even number. For simplicity we take $n=2$. The terms from

\begin{eqnarray}
\left[ b^+(\mathbf{p}_n) e^{i\varepsilon_{\mathbf{p}_n} \tau} - b(\mathbf{p}_n) e^{-i\varepsilon_{\mathbf{p}_n} \tau}\right]^5 ~,
\end{eqnarray} contributing the matrix element $\langle 0| \varPhi(t,\,\mathbf{x}) | \mathbf{p}_i \rangle$ are

\begin{eqnarray}
-\, e^{-i\varepsilon_{\mathbf{p}_n} \tau}\left[b(\mathbf{p}_n)b^+(\mathbf{p}_n)b(\mathbf{p}_n)b^+(\mathbf{p}_n)b(\mathbf{p}_n)
 \, + \right. ~~~~~\nonumber \\ b(\mathbf{p}_n)b^+(\mathbf{p}_n)b(\mathbf{p}_n)b(\mathbf{p}_n)b^+(\mathbf{p}_n) \, + ~~~\nonumber \\  b(\mathbf{p}_n)b(\mathbf{p}_n)b^+(\mathbf{p}_n)b^+(\mathbf{p}_n)b(\mathbf{p}_n) \, + ~\nonumber \\ \left. b(\mathbf{p}_n)b(\mathbf{p}_n)b^+(\mathbf{p}_n)b(\mathbf{p}_n)b^+(\mathbf{p}_n)\right]~. \end{eqnarray} Analogously, one finds that the terms form 
 
 \begin{eqnarray}
 \left[ b(\mathbf{p}_n) e^{-i\varepsilon_{\mathbf{p}_n} \tau} - b^+(\mathbf{p}_n) e^{i\varepsilon_{\mathbf{p}_n} \tau}\right]^5
 \end{eqnarray} contributing the matrix element $\langle 0| \varPhi(t,\,\mathbf{x}) | \mathbf{p}_i \rangle$ are   

\begin{eqnarray}
e^{-i\varepsilon_{\mathbf{p}_n} \tau}\left[b(\mathbf{p}_n)b^+(\mathbf{p}_n)b(\mathbf{p}_n)b^+(\mathbf{p}_n)b(\mathbf{p}_n)
 \, + \right. ~~~~~\nonumber \\ b(\mathbf{p}_n)b^+(\mathbf{p}_n)b(\mathbf{p}_n)b(\mathbf{p}_n)b^+(\mathbf{p}_n) \, + ~~~\nonumber \\  b(\mathbf{p}_n)b(\mathbf{p}_n)b^+(\mathbf{p}_n)b^+(\mathbf{p}_n)b(\mathbf{p}_n) \, + ~\nonumber \\ \left. b(\mathbf{p}_n)b(\mathbf{p}_n)b^+(\mathbf{p}_n)b(\mathbf{p}_n)b^+(\mathbf{p}_n)\right]~. \end{eqnarray} Hence, one finds

\begin{eqnarray}
\langle 0|\left[ b^+(\mathbf{p}_n) e^{i\varepsilon_{\mathbf{p}_n} \tau} - b(\mathbf{p}_n) e^{-i\varepsilon_{\mathbf{p}_n} \tau}\right]^5|\mathbf{p}_i \rangle &=& - 9 \delta_{in} e^{-i\varepsilon_{\mathbf{p}_n} \tau} ~, \nonumber \\
\langle 0|\left[ b(\mathbf{p}_n) e^{-i\varepsilon_{\mathbf{p}_n} \tau} - b^+(\mathbf{p}_n) e^{i\varepsilon_{\mathbf{p}_n} \tau}\right]^5|\mathbf{p}_i \rangle &=&  9 \delta_{in} e^{-i\varepsilon_{\mathbf{p}_n} \tau} ~.\nonumber 
\end{eqnarray} and correspondingly

\begin{eqnarray}\label{tavisupalitalghuri}
\langle 0| \varPhi(t,\,\mathbf{x}) | \mathbf{p}_i \rangle \,\propto\, e^{i(\mathbf{p}_i\mathbf{x} \,-\, \varepsilon_{\mathbf{p}_i} t)}\left( 1 \,-\,i \, \frac{\ss \mu^2 27\varepsilon_{\mathbf{p}_i}^3}{20}\,t \right) \,-\nonumber \\ \frac{\ss\mu^2 \, 27\, \varepsilon_{\mathbf{p}_i}^2}{40} \, e^{i(\mathbf{p}_i\mathbf{x} \,+\, \varepsilon_{\mathbf{p}_i} t)} \,+\, \frac{\ss\mu^2 \, 27\, \varepsilon_{\mathbf{p}_i}^2}{40} \, e^{-i(\mathbf{p}_i\mathbf{x} \,-\, \varepsilon_{\mathbf{p}_i} t)} ~.  ~~
\end{eqnarray}

\subsection{Corrections to the BH emission}

As it was discussed in section \ref{Abschnitt1}, if we subject the particles emitted by the BH to the modified dispersion relation \eqref{highdimmomentumrel} and retain in this equation only leading and subleading terms then BH entropy acquires logarithmic and inverse area corrections. Minimum-length deformed prescription applied at the second quantization level leads essentially to the same sort of corrections to the BH entropy \cite{Berger:2010pj}. Let us address this question in some detail.

The first term in the wave-function of a free particle \eqref{tavisupalitalghuri}           

\begin{eqnarray}e^{i(\mathbf{p}_i\mathbf{x} \,-\, \varepsilon_{\mathbf{p}_i} t)}\left( 1 \,-\,i \, \frac{27\ss\mu^2\varepsilon_{\mathbf{p}_i}^3}{20}\,t \right) \,\approx \, ~~~~~~~ \nonumber \\  e^{i(\mathbf{p}_i\mathbf{x} \,-\, \left[\varepsilon_{\mathbf{p}_i} \,+\,  1.35 \ss\mu^2 \varepsilon_{\mathbf{p}_i}^3 \right] t)} ~, \end{eqnarray} (to the first order in $\ss$) gives just the modified dispersion relation 

\[ \varepsilon_{\mathbf{p}_i} \,\rightarrow\, \varepsilon_{\mathbf{p}_i} \,+\,  1.35 \ss\mu^2 \varepsilon_{\mathbf{p}_i}^3 ~,\] where the energy scale $\mu$ is set by the BH emission temperature: $\mu=T$ \cite{Berger:2010pj}. It results in the logarithmic correction to the BH entropy, see section \ref{Abschnitt1}.

The second term in Eq.\eqref{tavisupalitalghuri} represents a reflected wave. In the context of the BH emission, it indicates the existence of the back scattered flux the rate of which is proportional to $\left|\ss\mu \varepsilon(\mathbf{p})^2\right|^2$, that is, to $l_F^8 T^8$. This flux increases with evaporating of the BH and thus tries to compensate the emission. It reproduces the inverse area correction to the entropy \cite{Berger:2010pj}. Namely, the standard Hawking temperature (for $n=2$) is defined as $T \propto \left( G_NM \right)^{-1/3}$. Hence, during the evaporation the BH mass changes as $dM \propto - dT/G_NT^4$ and for the mass increment due to back scattered flux one finds: $dM_+ \propto |\l_F^8T^8dM| \propto dT l_F^4T^4$. Using this equation and the formula $dS=dM/T$, one finds the entropy correction $\propto (l_F/r_g)^4$. 

The third term in Eq.\eqref{tavisupalitalghuri} could be interpreted as indicating the possibility of particle transition into an antiparticle. The discussion concerning this term can be found in \cite{Berger:2010pj}.

So far we have confined the application of the minimum-length deformed QM to the matter fields. But what if gravity (graviton field) is also subject to this sort of modification? Putting aside corrections arising at the second quantization level, one can address this question by estimating gravitational potential with the use of the modified propagator that follows from Eq.\eqref{wirkung}. The spherically symmetric gravitational field in $4+n$ space-time dimensions is described by the Schwarzschild-Tangherlini solution \cite{Tangherlini:1963bw, Myers:1986un}

\begin{eqnarray}  ds^2 \,=\, \left[ 1- \left(\frac{r_g}{r}\right)^{n+1} \right]dt^2 \,-\, ~~~~~~~~~~~~~~~~~~~ \nonumber \\  \left[ 1- \left(\frac{r_g}{r}\right)^{n+1} \right]^{-1}dr^2 \,-\, r^2 d\Omega^2_{n+2} ~,  \end{eqnarray} where $d\Omega^2_{n+2}$ is a line element of a $2+n$ dimensional unit sphere and the gravitational radius reads

\begin{eqnarray}\label{standardgradius}  r_g\left(M\right) \,=\, \left(\mathbb{G}_N M \right)^\frac{1}{n+1} \left[\frac{16 \pi }{(n+2)\text{Vol}\left(S^{n+2}\right)} \right]^\frac{1}{n+1} ~. \end{eqnarray}

\noindent Let us consider modified Schwarzschild-Tangherlini space-time 

\begin{eqnarray}  ds^2 \,=\, \left[ 1 \,-\, r_g^{n+1} V(r) \right]dt^2 \,-\, ~~~~~~~~~~~~~~~~~~~ \nonumber \\    \left[ 1 \,-\, r_g^{n+1}V(r) \right]^{-1}dr^2 \,-\, r^2 d\Omega^2_{n+2} ~, \end{eqnarray} where $r_g$ is given by Eq.\eqref{standardgradius} and $V(r)$ is calculated by the modified propagator with respect to Eq.\eqref{wirkung}

\begin{eqnarray}
V(r) \,=\, \frac{\text{Vol}\left(S^{2+n}\right)}{(2\pi)^{3+n}} \, \times ~~~~~~~~~~~~~~~~~~~~~~~~~~~~~~ \nonumber \\ \int\limits_{k^{2+n} <  \ss^{-1}}d^{3+n}k \, \frac{\left(1-\ss k^{2+n}\right)^{\frac{2}{1+n}}}{k^2} \, e^{i \mathbf{k}\mathbf{r}} ~,
\end{eqnarray} where now $\ss$ stands for (do not confuse with Eq.\eqref{ersteBezeichnung}) 

\begin{equation}\label{aghnishvna}
\frac{2\beta (1+n)l_F^{2+n}}{2+n} \, \equiv \, \ss ~. 
\end{equation} The potential $V(r)$ has the following generic properties. It is a monotonically decreasing function, finite at the origin with the vanishing derivative at this point, see Appendix. Its asymptotic behaviour when $r \rightarrow 0$ looks like

\begin{eqnarray}\label{potasyopaktsevamtsmandz}
V(r) \,=\, \mathcal{A} \,-\, \mathcal{B}r^2 \,+\, O\left(r^4\right)~,
\end{eqnarray} where $\mathcal{A}$ and $\mathcal{B}$ are positive quantities, see Appendix. Now the equation for the horizon looks like 

\begin{equation}\label{horizontisgantoleba}
\frac{1}{r^{n+1}_g\left(M\right)} \,=\, V(r) ~.
\end{equation} As $V(0)$ is a maximum of the potential, this equation does not have any solution for $M < M_{remnant}$, 

\begin{widetext}
\begin{equation}
\frac{1}{r^{n+1}_g\left(M_{remnant}\right)} \,=\, \frac{(n+2)\Gamma\left(\frac{3+n}{2}\right)}{32 \,\pi^\frac{n+5}{2}l_F^{2+n} M_{remnant}} \,=\, V(0) \,=\, \frac{\int\limits_0^1dq \, q^n \left(1-q^{2+n}\right)^{2/(1+n)}\int\limits_0^1 dt (1-t^2)^{n/2}}{2^n\sqrt{\pi} \, \Gamma\left(\frac{2+n}{2}\right)\Gamma\left(\frac{3+n}{2}\right)l_F^{1+n}}\, \left(\frac{2+n}{2\beta(1+n)}\right)^\frac{1+n}{2+n} ~,
\end{equation}\end{widetext} where we have used Eqs.(\ref{standardgradius}, \ref{aghnishvna}, \ref{horizontisgantoleba}). So, what we see is that as the BH evaporates down to the $M_{remnant}$ its horizon disappears and at the same time its surface gravity vanishes: $V'(0) = 0$ (see Eq.\eqref{potasyopaktsevamtsmandz}). That is, the Hawking temperature, which is proportional to the surface gravity becomes zero.

\subsection{Summary}

Let us briefly summarize the basic points of our discussion.

\noindent {\bf 1.} In deriving the higher-dimensional minimum-length modified QM we were guided by two recent papers \cite{Sen:2012dw, Bodendorfer:2013sja} demonstrating (in the framework of string theory and loop quantum gravity) that the logarithmic corrections to the BH entropy is universal in arbitrary space-time dimensions. The MUR we have found disagrees with the relation obtained in \cite{Scardigli:2003kr} but coincides with that one derived in \cite{Aurilia:2002aw} in somewhat different context. 

\noindent {\bf 2.} Using the Hilbert space representation for a relatively broad class of Planck-length deformed QM \cite{Maziashvili:2012dd} we consider minimum-length deformed QFT that follows from the higher-dimensional minimum-length deformed QM mentioned above. 

\noindent {\bf 3.} In discussing the minimum-length deformed QFT (both at first and second quantization levels) we restrict ourselves to the first order in deformation parameter (in this limit the corrections arising at the first and second quantization levels decouple). From the standpoint of the Einstein equations, up to this point we are just considering corrections to the matter fields. These corrections result in the logarithmic and inverse area corrections to the BH entropy. Thus providing a self-contained picture.

\noindent {\bf 4.1} General relativity viewed as a field theory in the Minkowskian background acquires corrections with respect to the minimum-length deformed QFT. Putting aside the corrections arising at the second quantization level, one can study the modified Schwarzschild-Tangherlini space-time by using the modified gravitational potential that comes from the minimum-length deformed QFT propagator. This way one finds a regular (de Sitter like) geometry near the origin. Indeed, that modified Schwarzschild-Tangherlini space-time is free of the curvature singularity at the origin because now the metric as well as its first and second derivatives do not diverge when $\mathbf{r} \rightarrow \mathbf{0}$ (see Eq.\eqref{potasyopaktsevamtsmandz}). On the other hand, the Schwarzschild-Tangherlini space-time modified this way shows up the zero-temperature BH remnants. The behaviour of the potential and Hawking temperature are plotted in Fig. 1 and Fig. 2, respectively. It should be remarked that the emission temperature in Fig. 2 vanishes when the BH horizon approaches zero.

\begin{figure}[h]
\includegraphics[width= 8cm, height= 7cm]{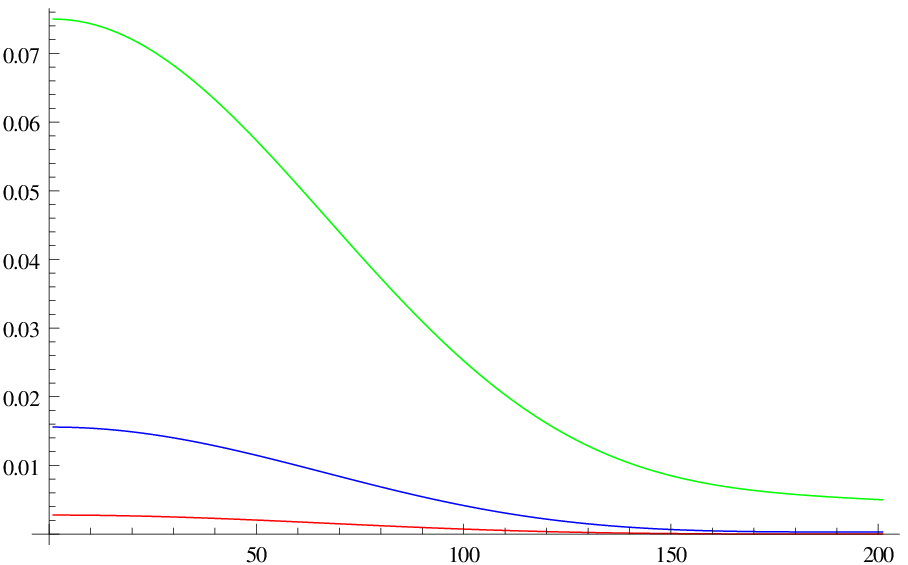}\\
\caption{Vertical axis: $V(r)\ss^{1/(2+n)}$ , horizontal axis: distance in units of $\ss^{1/(2+n)}$; $n=1$ green line, $n=2$ blue line, $n=3$ red line.}

\vspace{0.2cm}

\includegraphics[width= 8cm, height= 7cm]{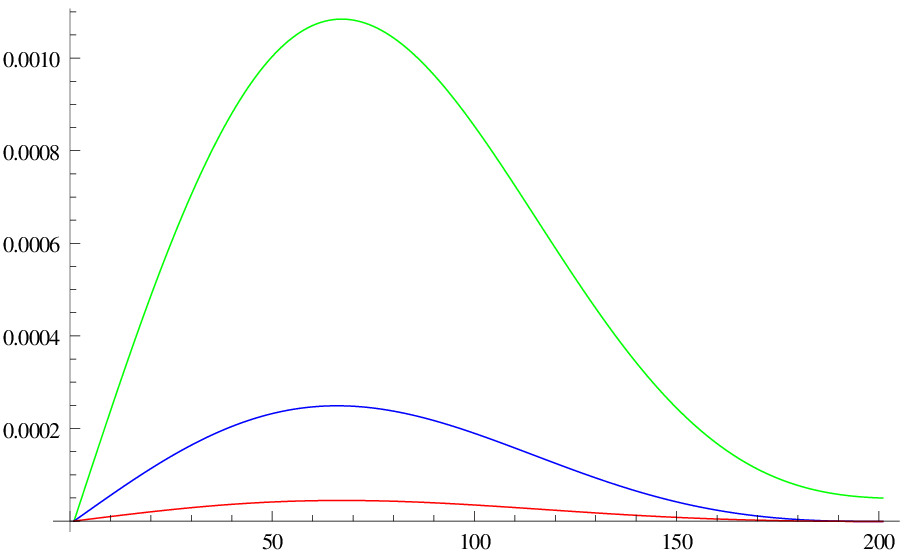}\\
\caption{Vertical axis: $T\ss^{1/(2+n)}/r_g^{n+1}$ , horizontal axis: distance in units of $\ss^{1/(2+n)}$; $n=1$ green line, $n=2$ blue line, $n=3$ red line.}

\end{figure}

\begin{figure}[h]
\includegraphics[width= 8cm, height= 7cm]{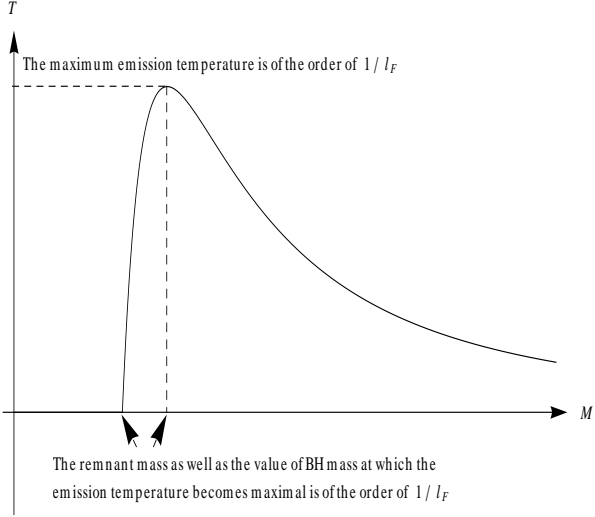}\\
\caption{Typical behaviour of the emission temperature as a function of the BH mass. The emission temperature reaches its maximum - of the order of $l_F^{-1}$, when BH evaporates down to the Planck mass, then it swiftly
drops to zero at $M_{remnant}$, which also is of the order of $l_F^{-1}$.
}
\label{potential}
\end{figure}

\noindent {\bf 4.2} In effect the approach we are pursuing starts from the modified Poisson equation: $\widehat{\mathbf{P}}^2V(r) \propto \delta_{\ss}^{(3+n)}(\mathbf{r})$, where the source energy density is given by the smeared-out ($3+n$ dimensional) $\delta$ function (in the limit $\ss \rightarrow 0$ one recovers the point-like source)  

\begin{eqnarray} \delta_{\ss}^{(3+n)}(\mathbf{r}) \,=\, \int\limits_{k^{2+n} <  \ss^{-1}}d^{3+n}k \,  e^{i \mathbf{k}\mathbf{r}} \,\propto \, \nonumber \\ \frac{1}{r^{(n+1)/2}} \int\limits_0^{\ss^{-1/(2+n)}} dk \, k^{(n+3)/2} J_{(n+1)/2}(kr) ~. \nonumber \end{eqnarray} So that BH we are discussing is surrounded by the matter. This sort of BHs are known as the "dirty" ones \cite{Visser:1992qh}. For our discussion we do nod need to address the generic picture of dirty BHs.          

Let us notice that one arrives at the regular BHs even just by considering the minimum-length deformed matter sector, when the smeared out source is taken in the framework of standard theory of gravity \cite{Nicolini:2005vd, Spallucci:2009zz, Casadio:2008qy, Nicolini:2008aj, Nicolini:2009gw}.

\noindent {\bf 4.3} Special attention has to be paid on the validity conditions of approximation assumed tacitly throughout the above discussion. We have taken gravitational field on the equal footing with the matter fields, that is, QFT picture for gravity is taken as a starting point. This means that the graviton field is defined as the difference between the full metric and its Minkowski background value. The calculations manifest that the gravity behaves as an asymptotically free interaction and, correspondingly, the radiative corrections close to the Planck scale can be safely ignored in this case \cite{Modesto:2010uh, Biswas:2011ar, Nicolini:2012eu}

\noindent {\bf 5.} Finally, let us remark that one can speculate about the possible observations of this sort of BH remnants in context of the large extra-dimensional models with low quantum gravity scale \cite{Antoniadis:1998ig, ArkaniHamed:1998rs, ArkaniHamed:1998nn, Silagadze:1999gr, Hossenfelder:2003doa, Hossenfelder:2007zz, Nicolini:2011nz, Bleicher:2010qr, Bleicher:2014laa}.

\subsection*{Appendix}

The integral determining the gravitational potential 
\begin{eqnarray}\label{appint}
\int\limits_{k^{2+n} < \, \ss^{-1}}d^{3+n}k \, \frac{\left(1-\ss k^{2+n}\right)^{\frac{2}{1+n}}}{k^2} \, e^{i \mathbf{k}\mathbf{r}}  ~, \end{eqnarray} for large values of $r$ is dominated by the wave-modes; $k\ll \ss^{-1/(2+n)}$, that is, in this limit the term $\ss k$ in the numerator can be neglected and respectively one recovers the standard result. To estimate its behaviour for small values of $r$, let us choose the axis $x_{3+n}$ along $\mathbf{k}$ and introduce spherical coordinates in the momentum space 

\begin{eqnarray}&& k_1 \,=\, k\sin\varphi \prod\limits_{j=1}^{n+1} \sin\theta_j ~,~~ k_2 \,=\, k\cos\varphi \prod\limits_{j=1}^{n+1} \sin\theta_j ~, \nonumber \\&& k_{i+2} \,=\, k\cos\theta_i \prod\limits_{j=i}^{n+1} \sin\theta_j~,~~ k_{3+n}  \,=\, k\cos\theta_{n+1}~, \nonumber \end{eqnarray} 

\noindent where $i = 1, \ldots , n~;~k \geq 0~,~ 0\leq \varphi < 2\pi~,~ 0 \leq \theta_j \leq \pi$. Thus, we get $\mathbf{k}\mathbf{r} = k x_{3+n} \cos\theta_{n+1}~,~d^{3+n}k = k^{2+n}\,dk\,d \varphi \prod\limits_{j=1}^{n+1} \sin^j\theta_j \, d\theta_j $, and the integral \eqref{appint} reduces to   

\begin{eqnarray}\label{shepasebadamateba}&&\text{Vol}\left(S^{n+1}\right)  \int\limits_0^{\ss^{-1/(2+n)}}  dk   k^n\left(1-\ss k^{2+n}\right)^{\frac{2}{1+n}} \times \nonumber \\&& \int\limits_0^{\pi} d\theta_{n+1} \, \sin^{n+1}\theta_{n+1}  \, e^{ik\, x_{3+n}\cos\theta_{n+1}} ~. \end{eqnarray}

Let us first consider the specific case $n=1$. Changing the variable $\cos\theta_2 = t$ one finds 

\begin{eqnarray}
\int\limits_0^{\pi} d\theta_{2} \, \sin^{2}\theta_{2}  e^{ik\, x_{4}\cos\theta_{2}}  = \int\limits_{-1}^1 dt \sqrt{1-t^2} \, \cos(k\, x_{4}t) = \nonumber\\ 2\int\limits_{0}^1 dt \sqrt{1-t^2} \, \cos(k\, x_{4}t)~, \nonumber
\end{eqnarray} and, correspondingly, the Eq.\eqref{shepasebadamateba} takes the form 

\begin{eqnarray}\label{ertidamganzpot}
8\pi \int\limits_{0}^1 dt \sqrt{1-t^2}  \int\limits_0^{\ss^{-1/3}}  dk \,  k\left(1-\ss k^{3}\right) \cos(k\, x_{4}t) ~.
\end{eqnarray} Performing the integrals

\begin{eqnarray}
\int\limits_0^{\ss^{-1/3}}  dk \,  k \cos(k\, x_{4}t) = \frac{d}{d(x_{4}t)}  \, \int\limits_0^{\ss^{-1/3}}  dk \,   \sin(k\, x_{4}t)  \nonumber \\ = \frac{d}{d(x_{4}t)} \,  \frac{1 - \cos\left( x_{4}t/\ss^{1/3}\right)}{x_{4}t} = \frac{\sin\left( x_{4}t/\ss^{1/(2+n)}\right)}{x_{4}t \ss^{1/3} } - \nonumber \\ \frac{1 - \cos\left( x_{4}t/\ss^{1/3}\right)}{x_{4}^2t^2}~, \nonumber \\
\int\limits_0^{\ss^{-1/3}}  dk \,  k^4 \cos(k\, x_{4}t) = \frac{d^4}{d(x_{4}t)^4}  \, \int\limits_0^{\ss^{-1/3}}  dk \,   \cos(k\, x_{4}t)  \nonumber \\ = \frac{d^4}{d(x_{4}t)^4} \, \frac{\sin\left( x_{4}t/\ss^{1/3}\right)}{x_{4}t}   =  \frac{\sin\left( x_{4}t/\ss^{1/3}\right)}{\ss^{4/3}x_{4}t} - \nonumber \\ \frac{4\sin\left( x_{4}t/\ss^{1/3}\right)}{\ss^{2/3}(x_{4}t)^3} - \frac{24\sin\left( x_{4}t/\ss^{1/3}\right)}{\ss^{1/3}(x_{4}t)^4} + \nonumber  \\ \frac{24\sin\left( x_{4}t/\ss^{1/3}\right)}{(x_{4}t)^5} ~, \nonumber 
\end{eqnarray} the final result reads

\begin{eqnarray}
V(r) \, = \,  \frac{1}{2\pi} \int\limits_{-1}^1 dt \sqrt{1-t^2} \left[\frac{4\beta^{1/3}\sin\left( rt/\ss^{1/3}\right)}{r^3t^3} \,+ \right. \nonumber \\   \frac{24\beta^{2/3}\sin\left( rt/\ss^{1/3}\right)}{r^4t^4}  \,-\,  \frac{1 - \cos\left( rt/\ss^{1/3}\right)}{r^2t^2} \,-\, \nonumber \\  \left. \frac{24\beta\sin\left( rt/\ss^{1/3}\right)}{r^5t^5} \right]~.~~
\end{eqnarray}

\noindent To find the asymptotic behaviour of the potential for $r \rightarrow 0$, one can immediately use the Eq.\eqref{ertidamganzpot}

\begin{eqnarray}
V(r) \, = \, \frac{3}{20\pi\ss^{2/3}} \int\limits_{-1}^1 dt \sqrt{1-t^2}   \,-\, \nonumber\\ \frac{3 r^2}{112\pi\ss^{4/3}} \int\limits_{-1}^1 dt \, t^2\sqrt{1-t^2}  \,+\, O\left(r^4\right) ~ .
\end{eqnarray}

\noindent From this expression it is immediate that $V(0)$ is finite and $V'(0)=0$. Now let us show that $V(r)$ is a monotonically decreasing function, that is, $V'(r) < 0$. From Eq.\eqref{ertidamganzpot} one finds   

\begin{eqnarray}
\frac{d}{dr}\int\limits_{0}^1 dt \sqrt{1-t^2}  \int\limits_0^{\ss^{-1/3}}  dk \,  k\left(1-\ss k^{3}\right) \cos(krt) = \nonumber \\ - \, \int\limits_{0}^1 dt \,t \sqrt{1-t^2}  \int\limits_0^{\ss^{-1/3}}  dk \,  k^2\left(1-\ss k^{3}\right) \sin(krt) = \nonumber \\  - \,\frac{1}{r^3} \int\limits_{0}^1 dt \,t \sqrt{1-t^2}  \int\limits_0^{r/\ss^{1/3}}  d\widetilde{k} \,  \widetilde{k}^2\left(1-\frac{\ss \widetilde{k}^{3}}{r^3}\right) \sin\left(\widetilde{k}t\right) < 0~; \nonumber 
\end{eqnarray} then from the statement 

\[ \int\limits_0^{a} f(z)\sin(z) \,>\, 0~,  \] whenever $f(z)$ is a positive and monotonically decreasing function, readily follows what we wanted to show.

Now let us address the general case. Denoting $\cos\theta_{n+1} = t$ one finds 

\begin{eqnarray}  \int\limits_0^{\pi} d\theta_{n+1} \, \sin^{n+1}\theta_{n+1}  \, e^{ik\, x_{3+n}\cos\theta_{n+1}} = \nonumber \\ 2 \int\limits_0^1 dt \left(1 \,-\, t^2 \right)^{n/2}\cos\left(kx_{3+n}t\right) ~, \nonumber \end{eqnarray} and correspondingly the potential takes the form

\begin{eqnarray}&&V(r) = \frac{2\text{Vol}\left(S^{n+1}\right)\text{Vol}\left(S^{n+2}\right)}{(2\pi)^{3+n}}  \int\limits_0^{\ss^{-1/(2+n)}}  dk   k^n \,\times \nonumber \\&& \left(1-\ss k^{2+n}\right)^{\frac{2}{1+n}}  \int\limits_0^1 dt \left(1 \,-\, t^2 \right)^{n/2}\cos\left(krt\right)  ~. \end{eqnarray} Its asymptotic behaviour for $r \rightarrow 0$ can readily be found by expanding the $\cos\left(krt\right)$ into the Taylor series

\begin{eqnarray}
V(r) = \frac{2\text{Vol}\left(S^{n+1}\right)\text{Vol}\left(S^{n+2}\right)}{(2\pi)^{3+n}} \left[\int\limits_0^{\ss^{-1/(2+n)}}  dk   k^n \,\times \nonumber \right.\\  \left(1-\ss k^{2+n}\right)^{\frac{2}{1+n}}  \int\limits_0^1 dt \left(1 \,-\, t^2 \right)^{n/2} - \frac{r^2}{2} \int\limits_0^{\ss^{-1/(2+n)}}  dk    \,\times \nonumber \\ \left. k^{n+2}\left(1-\ss k^{2+n}\right)^{\frac{2}{1+n}}  \int\limits_0^1 dt \, t^2\left(1 \,-\, t^2 \right)^{n/2} \,+\, O\left(r^4\right)\right] ~ .~~
\end{eqnarray} It is evident from this expression that $V(0)$ is finite, $V'(0)=0$ and $V'(r) < 0$ for $r \rightarrow 0$. In general, the statement $V'(r) < 0$ for $r>0$ can be proved much in the same way as it was done for $n=1$.

\acknowledgments

The substantial part of this work was done while I was a visitor at the Frankfurt Institute for Advanced Studies; I am greatly indebted to Marcus Bleicher and Piero Nicolini for their invitation and hospitality. This research was partially supported by the Shota Rustaveli National Science Foundation under contract number 31/89 and DAAD research fellowship for university teachers and researchers.

\end{document}